\documentclass[twocolumn,superscriptaddress,prl,showpacs]{revtex4}
\usepackage[letterpaper,left=0.7in,right=0.7in,top=0.90in,bottom=0.8in]{geometry}
\usepackage{amssymb}
\usepackage{amsmath}
\usepackage{graphicx}
\usepackage{subfigure}

\begin{document}

\title{
Strength of the symmetry spin filtering effect in magnetic tunnel junctions
}

\author{Sergey V. Faleev}
\email{svfaleev@us.ibm.com}
\affiliation{IBM Almaden Research Center, 650 Harry Road, San Jose, California 95120, USA}
\author{Stuart S. P. Parkin}

\affiliation{IBM Almaden Research Center, 650 Harry Road, San Jose, California 95120, USA}
\author{Oleg N. Mryasov}
\affiliation{Physics and Astronomy, University of Alabama, Tuscaloosa, AL, 35487, USA}

\date{\today}

\begin{abstract}
Strength of the the symmetry spin filtering effect (as defined by the asymptotic behavior of the tunneling magnetoresistance  (TMR) at large barrier thicknesses induced by this effect) is studied for the Fe/MgO/Fe magnetic tunnel junctions (MTJ).
Based on the analysis of the band structure of bulk Fe and complex band structure of MgO  we predict \emph{native} for the symmetry spin filtering effect linear increase of the TMR in Fe/MgO/Fe MTJ with increasing number of MgO layers, $N$. \emph{Ab initio} calculations of transmission functions performed for the Fe/MgO/Fe MTJ confirm our theoretical predictions
for the strength of  the symmetry spin filtering effect in broad range of energies and $N$. Our calculations  also show that the \emph{combination} of the symmetry spin filtering effect and small surface transmission function in minority spin channel at the Fe/MgO interface is responsible for large $TMR>10,000\%$ predicted for Fe/MgO/Fe MTJ  for $N \geqslant 8$.
Proposed analysis of the strength of the symmetry filtering effect derived from the band structure of bulk electrode material could serve as a tool for quick material discovery search of suitable electrodes in context of emerging technologies that require high TMR.
\end{abstract}
\maketitle


Theoretical prediction of high TMR in Fe/MgO/Fe MTJ due to so-called symmetry spin filtering mechanism \cite{Butler01,Mathon01}   and its quick experimental verification \cite{Parkin04,Yuasa04} revolutionized the hard disk drive (HDD) industry during the last decade. But despite a lot of theoretical and experimental attention to this MTJ dependence  of the TMR on the number of MgO layers, $N$, arising from native symmetry spin filtering mechanism is still not fully understood and somewhat controversial \cite{Parkin04}.
Theoretical calculations based on the density functional theory (DFT) predict that in ideal Fe/MgO/Fe  junction  TMR should increase very fast with increasing $N$. More specifically, TMR is predicted to change by as much as two orders of magnitude when $N$ changes form 4 to 12 \cite{Butler01,Belashchenko05}. In contrast, experimental measurements show that TMR does not depend much on the thickness of MgO \cite{Parkin04,Yuasa04}.


We explain the controversy between theoretical calculations and experimental results  by the fact that fast increase of the TMR predicted previously \cite{Butler01,Belashchenko05} is a consequence of the  contribution to transmission function from the interface resonance states (that exist in minority Fe channel in a very narrow energy window near the Fermi energy, $E_F$ \cite{Butler01,Belashchenko05,Rungger09,Tiusan04,Feng09,Faleev12}), while native symmetry spin filtering effect, in general, leads to modest linear increase of the TMR with $N$ in the asymptotic limit   $N \to \infty$.
Thus, absence of strong dependence  of measured TMR on $N$ could be explained by the interface roughness that destroys the interface resonance states (IRS).

In this paper we also describe features of the band structure of \emph{bulk} electrode material that give more stronger $\propto N^2$ dependence  of the TMR on $N$ in the limit   $N \to \infty$. Proposed below analysis of the strength of the symmetry filtering effect based on the bulk band structure of candidate electrode material could serve as a tool for quick material discovery search of suitable electrodes in context of emerging technologies that require high TMR. As an example of such technology that critically depends on discovery of novel MTJs with high TMR  we mention spin-transfer torque   magnetoresistive random-access memory (STT-MRAM) technology (that has a potential to become an 'universal memory' \cite{Akerman05}) where the pool of candidate electrode materials includes several hundreds  Heusler allays, magnetic multilayers, etc.


For the Fe/MgO/Fe MTJ  with sufficiently large MgO thickness the transmission function for electrons with in-plane wave vector $\mathbf{k}$ and energy $E$ inside the MgO band gap is determined by single surviving evanescent state inside the MgO barrier at this $\mathbf{k}$ and $E$, $\psi_e(\mathbf{k},E)$,  that has smallest attenuation constant, $\gamma(\mathbf{k},E)$.  Transmission function in the limit   $N \to \infty$ is given by \cite{Belashchenko04}
\begin{equation}
T_{\sigma \sigma'}(\mathbf{k},E)=t_{\sigma \mathbf{k} E} \times
e^{-\gamma(\mathbf{k},E) N} \times t_{\sigma' \mathbf{k} E} \ , \label{eqTT}
\end{equation}
where subindexes $\sigma$ and $\sigma'$ describe the spin channel of the left and right electrodes, correspondingly. (We use notations where $\sigma$ takes two values, $u$ and $d$  (short for "up" and "down") for majority and minority spin channel, correspondingly. Thus, $T_{uu}$ and $T_{dd}$ are majority-majority and minority-minority transmission in parallel configuration (PC) of the electrodes, and $T_{ud}$ and $T_{du}$ are majority-minority and minority-majority transmission in antiparallel configuration (APC) of the electrodes.) The coefficient $t_{\sigma \mathbf{k} E}$ in  (\ref{eqTT}) is the so-called surface transmission function (STF) defined for each electrode separately (in the case of different electrodes) by solution of the scattering problem  at  the electrode-barrier interface
\begin{equation}
t_{\sigma \mathbf{k} E} = \sum_p |B_e/A_p|^2 \ . \label{eq2}
\end{equation}
Here summation is taken over all eigenstates $p$ of the electrode with given $\sigma$, $\mathbf{k}$ and E, $A_p$ is amplitude of the eigenstate $p$ incoming from the electrode and $B_e$ is corresponding amplitude of the scattering  wavefunction inside the barrier taken at the reference plane - plane located at sufficient distance from the interface where  scattering wavefunctions for all $p$ are already indistinguishable from surviving evanescent state $\psi_e(\mathbf{k},E)$. Strictly speaking, with such definition of the $t_{\sigma \mathbf{k} E}$,  $N$ in Eq. (\ref{eqTT}) is the number of MgO layers between reference planes corresponding to the two electrode-barrier interfaces, but we will use total number of MgO layers, $N$, in Eq. (\ref{eqTT}) assuming proper re-definition of the $t_{\sigma \mathbf{k} E}$. In general, for different electrodes, $t_{\sigma \mathbf{k} E}$ should also have the electrode index (left or right), but for Fe/MgO/Fe MTJ two electrodes are the same, so the notation $t_{\sigma \mathbf{k} E}$ without reference to the left or right electrode is used in (\ref{eqTT}).

Total transmission of the MTJ is given by the $\mathbf{k}$-integral over the 2D surface Brillouin zone (SBZ)
\begin{equation}
T_{\sigma \sigma'}(E)=\int \frac{d^2\mathbf{k}A}{(2\pi)^2} T_{\sigma \sigma'}(\mathbf{k},E)
= \sum_{\mathbf{k}} T_{\sigma \sigma'}(\mathbf{k},E)
 \ , \label{eq1}
\end{equation}
where $A$ is the in-plane cross-sectional area of the device.

We emphasize two important features of the Eq. (\ref{eqTT}) for transmission function: (1) due to the flux conservation the same STF $t_{\sigma \mathbf{k} E}$ describes two different processes - transmission from the electrode to the barrier and transmission from the barrier to the electrode, and (2) STF of two electrodes are independent  from each other (electrodes are decoupled). One nontrivial consequence of decoupling of the two electrodes and transmission through single channel inside the barrier as described by Eq. (\ref{eqTT}) is that in the limit  $N \to \infty$ transmission in APC  can be expressed in terms of transmission functions for majority and minority electrons in PC, $\lim_{N \to \infty}T'_{ud}(E) = T_{ud}(E)$, where
\begin{equation}
T'_{ud}(E) = \sum_{\mathbf{k}} [T_{uu}(\mathbf{k},E) \times T_{dd}(\mathbf{k},E) ]^{1/2} \ . \label{TAP1}
\end{equation}

In order to determine at what barrier thickness the asymptotic expression (\ref{eqTT}) becomes valid
in the case of Fe/MgO/Fe MTJ we performed calculations of transmission functions $T_{ud}(E)$ and $T'_{ud}(E)$  for different $N$ using an \emph{ab-initio} tight-binding linear muffin-tin orbitals (TB-LMTO) methods in its atomic spheres approximation (ASA) [\onlinecite{Schilfgaarde98,Turek97,Faleev05}]. Results of these calculations for $N=4$, 6, 8, 10, and 12 are shown on Fig. 1(a). One can see that  $T_{ud}(E)$ defined by Eq. (\ref{eq1}) and $T'_{ud}(E)$ defined by Eq. (\ref{TAP1})  indeed are very close to each other even for $N=4$. For larger $N$ agreement between $T_{ud}(E)$ and  $T'_{ud}(E)$ becomes better and at $N=12$ $T_{ud}(E)$ and $T'_{ud}(E)$  are almost indistinguishable. We conclude that the asymptotic behaviour described by Eq. (\ref{eqTT}) is reached for the Fe/MgO/Fe MTJ starting already with $N=4$.

\begin{figure}[t]
\includegraphics*[trim={0.3cm 16.0cm 1.4cm 2.4cm},clip,width=8.5cm]{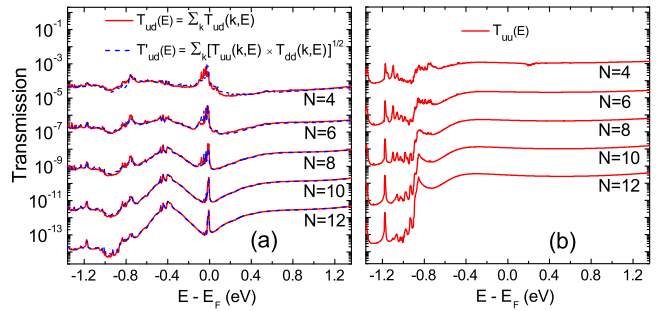}
\caption{(color online). (a) Comparison of two expressions for the APC transmission, $T_{ud}(E) = \sum_{\mathbf{k}} T_{ud}(\mathbf{k},E)$ (red lines), and  $T'_{ud}(E)= \sum_{\mathbf{k}} [T_{uu}(\mathbf{k},E) \times T_{dd}(\mathbf{k},E)]^{1/2} $ (blue lines) for the Fe/MgO/Fe MTJ with number of MgO layers $N=4$, 6, 8, 10, and 12. (b) Majority-majority transmission in PC, $T_{uu}(E)$,  for the Fe/MgO/Fe MTJ with different $N$. }
\label{fig1}
\end{figure}


Evanescent state of MgO with smallest attenuation constant is the state  $\psi_e(\mathbf{k},E)$ with $\mathbf{k}=0$ \cite{Butler01}.  This state has $\Delta_1$ symmetry (function $\psi_e(0,E)$ stays invariant with respect to the square-symmetry transformations of $(x,y)$ coordinates). For small $\mathbf{k}$  the  attenuation constant could be approximated as
\begin{equation}
\gamma_{\mathbf{k},E} = \gamma(0,E) + \alpha k^2 \ . \label{att}
\end{equation}

It is well known \cite{Butler01} that Fe has $\Delta_1$-symmetry  \emph{majority} electron band  along the $\Gamma-$H  line in 3D BZ ($\Gamma-$H line in 3D BZ corresponds to the $\mathbf{k}=0$ in 2D SBZ) with energies near $E_F$ (see Fig \ref{fig4}(b)). This  $\Delta_1$-symmetry  state could couple with the $\Delta_1$-symmetry evanescent state of MgO, so the majority STF  $t_{u, \mathbf{k}=0, E}$ is not zero. On the other hand, Fe does not have \emph{minority} electron bands with $\Delta_1$-symmetry along the $\Gamma-$H line with energies near $E_F$ (see  Fig \ref{fig4}(c)). Thus, minority STF $t_{d, \mathbf{k}=0, E} = 0$ (for $\mathbf{k}=0$ the overlap integral at the Fe/MgO interface between Fe minority states with symmetries other then  $\Delta_1$-symmetry and the evanescent state of MgO with $\Delta_1$-symmetry is zero by the symmetry).

The $\Delta_1$-symmetry evanescent state $\psi_e(0,E)$ of MgO mostly consist of $s$-orbitals of Mg and $p_z$ orbitals of O. In the spirit of the perturbation theory, for small but non-zero $\mathbf{k}$ the eigenstate function $\psi_e(\mathbf{k},E)$ could be represented as $\psi_e(0,E)$ plus some small terms proportional to $\mathbf{k}$ (we consider here changes in the orbital composition of the  eigenstate function inside the unit cell and factor out the global unit cell to unit cell translation factor $\exp{(\mathbf{kr})}$). In the linear over $\mathbf{k}$ approximation the $p_z$ orbital of O that is aligned along \emph{z}-axis will rotate to small angle proportional to $|k|$, so  $\psi_e(\mathbf{k},E)$ will include $p_x k_x$ and $p_y k_y$ terms where $p_x$ and $p_y$ are the \emph{p}-orbitals of the O atom.
At the Fe/MgO interface the $p_x$ and $p_y$ orbitals of  O have non-zero overlap integral with two minority Fe $\psi^{\Delta_5}_d(\mathbf{k},k_z)$  eigenstates composed primary of $d_{xz}$ and $d_{yz}$ orbitals of Fe atoms (two $\psi^{\Delta_5}_d(\mathbf{k},k_z)$ Fe minority bands are the extensions to non-zero $\mathbf{k}$ of two degenerate $\Delta_5$-symmetry bands $\psi^{\Delta_5}_d(0,k_z)$ that exists at energies near $E_F$  along the $\Gamma-$H line ($\mathbf{k}=0$ in SBZ), see  Fig 4(c)).

Analogously, these two  minority Fe eigenstates $\psi^{\Delta_5}_d(\mathbf{k},k_z)$  for small but non-zero $\mathbf{k}$ could be represented as $\psi^{\Delta_5}_d(0,k_z)$ plus small terms proportional to $\mathbf{k}$. In the linear over  $\mathbf{k}$ approximation the $d_{xz}$ and $d_{yz}$ orbitals of   two $\psi^{\Delta_5}_d(0,k_z)$ eigenfunctions will   rotate   by small angles in order to accommodate new propagation direction, $(k_x,k_y,k_z)$ instead of $(0,0,k_z)$, of the $\psi^{\Delta_5}_d(\mathbf{k},k_z)$ eigenfunctions (here $k_z$ is the wave-vector that corresponds to the energy $E$ of the $\Delta_5$-symmetry band at $\mathbf{k}=0$). (Note that rotation of the $d_{xz}$ and $d_{yz}$ orbitals of two $\psi^{\Delta_5}_d$ eigenstates in accordance with the propagation direction is consistent with the fact that propagating along rotated by $90^o$ direction, e.g. along the $x$-axis, wave functions $\psi^{\Delta_5}_d$ are composed of rotated by $90^o$  $d_{xz}$ and $d_{xy}$ orbitals.) It is straightforward to show that $d_{xz}$ orbital rotated around $y$-axis on  small angle $\theta=k_x/k_z$ will have $d_{zz}$ component with weight proportional to $k_x$. Similarly, $d_{yz}$ orbital rotated around $x$-axis on  small angle $\theta=k_y/k_z$ will have $d_{zz}$ component with weight proportional to $k_y$. This $d_{zz}$ component is invariant with respect to rotation around \emph{z}-axis and, at the Fe/MgO interface, it has non-zero overlap integral with $s$-orbitals of Mg and $p_z$-orbitals of O of the $\psi_e(\mathbf{k},E)$ MgO eigenstate .

Both described above contributions to the overlap integral at the Fe/MgO interface between either of the two $\psi^{\Delta_5}_d(\mathbf{k},k_z)$ eigenstates and $\psi_e(\mathbf{k},E)$ eigenstate  are proportional to $\mathbf{k}$ for small $\mathbf{k}$. Thus, the scattering amplitude $B_e$ inside the the barrier  originated from  incoming $\psi^{\Delta_5}_d(\mathbf{k},k_z)$ is proportional to $\mathbf{k}$. Since the expression for the $t_{\sigma,\mathbf{k}, E}$ (\ref{eq2}) includes the square of the scattering amplitude, $|B_e|^2$, the $t_{d,\mathbf{k}, E}$ is proportional to the $k^2$ for small $\mathbf{k}$.

Using Eq. (\ref{eqTT}) with above estimations for majority and minority STFs, $t_{u,\mathbf{k}, E} \propto 1$ and $t_{d,\mathbf{k}, E} \propto k^2$, we can write expressions for $\mathbf{k}$-resolved transmission functions in the limit of small $\mathbf{k}$:
\begin{eqnarray}
T_{uu}(\mathbf{k},E) &=&  A_{uu} e^{-(\gamma(0,E) + \alpha k^2)N}
\  \label{T.1} \\
T_{ud}(\mathbf{k},E) &=&  A_{uu}f_{ud}(\mathbf{k}/|k|) k^2  e^{-(\gamma(0,E) + \alpha k^2)N}
\    \label{T.2} \\
T_{dd}(\mathbf{k},E) &=&  A_{uu}f^2_{ud}(\mathbf{k}/|k|) k^4  e^{-(\gamma(0,E) + \alpha k^2)N}  \label{T.3}
\
\end{eqnarray}
Here $A_{uu}$ is a constant (for fixed energy), and $f_{ud}(\mathbf{k}/|k|)$, in general, is functions of the $\mathbf{k}$-direction, $\mathbf{k}/|k|$. Note, that in the Eq. (\ref{T.3}) square of the function $f_{ud}(\mathbf{k}/|k|)$ is used, as prescribed by the Eq. (\ref{eqTT}) that demands the equality $T_{uu}(\mathbf{k},E) \times T_{dd}(\mathbf{k},E) = T^2_{ud}(\mathbf{k},E)$.

The large-$N$ asymptotic behavior of transmission functions for  energies near $E_F$ could be easily obtained now by performing the $\mathbf{k}$ integration in Eq. (\ref{eq1}):
\begin{eqnarray}
T_{uu}(E) &\propto & \int d^2\mathbf{k}  e^{-(\gamma(0,E) + \alpha k^2)N} \propto \frac{ e^{-\gamma(0,E)N}}{N}
\  \label{T3.1} \\
T_{ud}(E) &\propto & \int d^2\mathbf{k}  k^2 e^{-(\gamma(0,E) + \alpha k^2)N} \propto \frac{ e^{-\gamma(0,E)N}}{N^2}
\    \label{T3.2} \\
T_{dd}(E) &\propto & \int d^2\mathbf{k}  k^4 e^{-(\gamma(0,E) + \alpha k^2)N}   \propto \frac{ e^{-\gamma(0,E)N}}{N^3} \label{T3.3}
\
\end{eqnarray}
From Eqs. (\ref{T3.1}-\ref{T3.3}) we can finally derive the large-$N$ asymptotic behaviour of TMR for energies near $E_F$
\begin{eqnarray}
TMR = \frac{T_{uu} + T_{dd} - 2 T_{ud}}{2 T_{ud}} = C N + D + O(1/N) \ .
\label{TMR}
\end{eqnarray}
Eq. (\ref{TMR}) gives \emph{native} asymptotic  behavior of the TMR enhanced due to the symmetry spin filtering effect (in other words, the strength of the effect) in the Fe/MgO/Fe MTJ.

The $\mathbf{k}$-resolved transmission $T_{ud}(\mathbf{k},E)$ is not equal to zero at $\mathbf{k}=0$
if contributions of the higher-order channels inside the MgO barrier with attenuation constants $\gamma'(\mathbf{k},E) > \gamma(\mathbf{k},E)$ (and symmetry other than the $\Delta_1$-symmetry at $\mathbf{k}=0$) are taken into account. Such contributions to $T_{ud}(\mathbf{k},E)$ at $\mathbf{k}=0$  were used to explain the nature of the symmetry spin filtering effect in existing literature \cite{Butler01}. We note, however, that contribution to the $\mathbf{k}$-integrated transmission $T_{ud}(E)$ from such terms is proportional to  $\exp{(-\gamma'(0,E)N)}$ and negligible compared to Eq. (\ref{T3.2}) in the limit of large $N$.

Authors of recent work \cite{Heiliger08} correctly found that $T_{uu}(\mathbf{k},E)$ and $T_{ud}(\mathbf{k},E)$ transmissions in the Fe/MgO/Fe MTJ have the same decay rate  for sufficiently thick  MgO barrier and  obtained correct linear increase of TMR with $N$ for $15<N<30$ at the Fermi energy (the effect of the IRS has not been discussed in the paper). However, they  did not consider the effect of the difference of the majority and minority STFs at  $|k|\sim 0$  on the $\mathbf{k}$-integrated transmissions and concluded, in disagreement with their own numerical results, that in the limit $N \to \infty$ TMR will eventually saturate to a constant.
Linear with $N$  behavior of TMR for $10<N<20$ in the Fe/MgO/Fe MTJ has been also obtained within a tight-binding framework   in \cite{Mathon01} and later reproduced in \cite{Mathon06} where effect of the chemical disorder at the interface was studied.

The asymptotical behavior of the $TMR(E)$ described by Eq. (\ref{TMR}) exists in the energy window from $E_F-0.4$ eV to $E_F+1.4$ eV. For $E > E_F+1.4$ eV the $\Delta_1$-symmetry band appears in minority channel at the $\Gamma-$H line (see Fig 4 (c)), so $TMR(E)$ will decrease to $TMR\propto 1$ at $E > E_F+1.4$ eV. For $E < E_F-1.0$ eV the $\Delta_1$-symmetry band disappears in  majority channel (see Fig 4 (b)), so $TMR(E)$ again will decrease to $TMR(E)\propto 1$ at $E < E_F-1.0$ eV. For $E < E_F-0.4$ eV the $\Delta_5$-symmetry band disappears in minority channel (see Fig 4 (c)). Thus, in the range of the energies $E_F - 1.0$ eV $ < E < E_F-0.4$ eV the $\Delta_1$-symmetry band still exists in majority channel and only $\Delta_2$-symmetry band exists in minority channel along the $\Gamma-$H line (see Fig 4(b,c)).

In the spirit of the perturbation theory, the minority Fe eigenstate $\psi^{\Delta_2}_d(\mathbf{k},k_z)$  for small but non-zero $\mathbf{k}$ could be represented as $\psi^{\Delta_2}_d(0,k_z)$ plus small terms proportional to $\mathbf{k}$ (the $\psi^{\Delta_2}_d(\mathbf{k},k_z)$ Fe minority band is the extension to non-zero $\mathbf{k}$ of the $\Delta_2$-symmetry band $\psi^{\Delta_2}_d(0,k_z)$   shown on Fig \ref{fig4}(c) along the $\Gamma-$H line). The eigenstate $\psi^{\Delta_2}_d(0,k_z)$ is composed primary of the $d_{x^2-y^2}$ orbitals of Fe. In the linear over  $\mathbf{k}$ approximation the $d_{x^2-y^2}$ orbital  will rotate  to small angle in order to accommodate new propagation direction $(k_x,k_y,k_z)$ instead of $(0,0,k_z)$ (here $k_z$ is the wave-vector that corresponds to energy $E$ for the $\Delta_2$-symmetry band at $\mathbf{k}=0$). It is straightforward to show that $d_{x^2-y^2}$ orbital rotated around $x$-axis or $y$-axis on small angle proportional to $|k|$ in the linear over $|k|$ approximation will have $d_{xz}$ and $d_{yz}$ components, but not $d_{zz}$ component. The $d_{zz}$ component appears only in second order over $|k|$.
Schematically, the expansion over orders of $|k|$ of the orbital composition of the $\psi^{\Delta_2}_d(\mathbf{k},k_z)$ Fe eigenstate and   $\psi_e(\mathbf{k},E)$ MgO eigenstate can be presented as:
\begin{eqnarray}
\psi^{\Delta_2}_u(\mathbf{k},k_z) &\sim & d_{x^2-y^2} + |k|(d_{xz} + d_{yz}) + |k|^2(d_{zz}+..) \nonumber \\
\psi_e(\mathbf{k},E) &\sim & (s+p_z) +  |k|(p_x+p_y) + O(|k|^2) \ .
\label{schem}
\end{eqnarray}
One can wee the the overlap integral at the Fe/MgO interface between the $\psi^{\Delta_2}_d(\mathbf{k},k_z)$  and   $\psi_e(\mathbf{k},E)$ eigenfunctions is non-zero only in the second order over $|k|$.  Thus, the scattering amplitude $B_e$ inside the the barrier  originated from  incoming $\psi^{\Delta_2}_d(\mathbf{k},k_z)$ eigenstate is proportional to $|k|^2$ and the STF $t_{d,\mathbf{k}, E}$ is proportional to  $k^4$ for small $\mathbf{k}$ in the energy window $E_F - 1.0$ eV $ < E < E_F-0.4$ eV. In this energy window the $T_{uu}(\mathbf{k},E)$ is still given by Eq. (\ref{T.1}), while $T_{ud}(\mathbf{k},E)$ and $T_{dd}(\mathbf{k},E)$ will contain more orders of $|k|^2$ at small $|k|$:
\begin{eqnarray}
T_{ud}(\mathbf{k},E) &=&  A_{uu}f_{ud}(\mathbf{k}/|k|) k^4  e^{-(\gamma(0,E) + \alpha k^2)N}
\    \label{T.4} \\
T_{dd}(\mathbf{k},E) &=&  A_{uu}f^2_{ud}(\mathbf{k}/|k|) k^8  e^{-(\gamma(0,E) + \alpha k^2)N}  \label{T.5}
\
\end{eqnarray}

Performing the $\mathbf{k}$ integration (\ref{eq1}) we can obtain the asymptotic expression for $TMR(E)$ at energies $E_F - 1.0$ eV $ < E < E_F-0.4$ eV:
\begin{eqnarray}
TMR(E) \propto N^2  \ ,
\label{TMR2}
\end{eqnarray}
which, at practical MgO thicknesses, $N \sim 10$, gives an additional order of magnitude enhancement of the TMR compared to the $TMR\propto N$ case described by Eq. (\ref{TMR}).

In order to confirm theoretical formulas derived above  we performed \emph{ab initio} calculations of transmission functions for the Fe/MgO/Fe MTJ with $N=4,6,8,10,$ and $12$ by using the TB-LMTO-ASA Green's function approach [\onlinecite{Schilfgaarde98,Turek97,Faleev05}]. We used relaxed nuclear coordinates of the Fe/MgO interface from Ref. [\onlinecite{Worthmann04}].

On Fig 2 (a) we show attenuation constant $\gamma(0,E)$ estimated from Eq. (\ref{T3.1})   using   $T_{uu}(E)$ for $N=10$ and $12$:
\begin{eqnarray}
\gamma(0,E) = \frac{1}{2} \ln{\left(   \frac{10T_{uu}(E,N=10)}{12T_{uu}(E,N=12)}    \right)}
\label{gamma}
\end{eqnarray}
In order to verify convergence of calculated $\gamma(0,E)$ with respect to $N$ we plotted  the product $T_{uu}(E)N \exp{[\gamma(0,E)N]}$ for $N=6,8,10$ and $12$ on Fig. 2(b). As can be seen  the curves for $N=8,10$ and $12$ are indistinguishable on the figure confirming both validity of the asymptotic formula (\ref{T3.1}) and convergence of calculated $\gamma(0,E)$ with respect to $N$ for broad range of energies. Decline of the $\gamma(0,E)$ at  $E = E_F - 0.85$ eV could be explained by approaching the edge of the $\Delta_1$-symmetry  majority  band that occurs at the energy slightly below $E = E_F - 0.85$ eV (see Fig 4(b)).

\begin{figure}[h]
\includegraphics*[trim={0.4cm 8.4cm 0.5cm 3.0cm},clip,width=8.5cm]{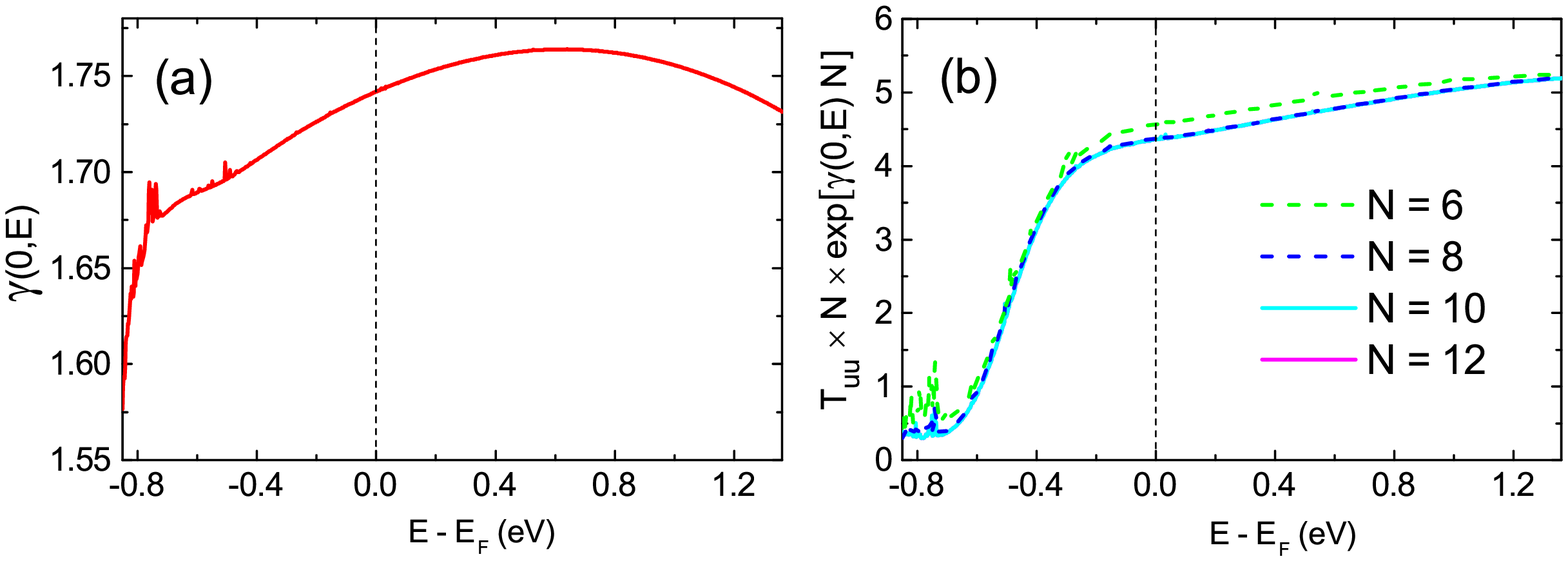}
\caption{(color online). Attenuation constant $\gamma(0,E)$ estimated from Eq.  (\ref{gamma}). (b)   $T_{uu}(E)N \exp{[\gamma(0,E)N]}$ calculated for $N=6,8,10$ and $12$. Curves with $N=8,10$ and $12$ are indistinguishable on the figure.
}
\label{fig2}
\end{figure}

The $\mathbf{k}$-resolved transmission functions $T_{uu}(\mathbf{k},E)$, $T_{ud}(\mathbf{k},E)$, and $T_{dd}(\mathbf{k},E)$ calculated for the Fe/MgO/Fe MTJ with $N=10$ for 6 energy points $E-E_F=-0.8, -0.4, 0, 0.05, 0.4$ and $0.8$ eV are presented on 6 panels of Fig. 3 as functions of the absolute value of the wave-vector $|k|$ (shown in units of $2\pi/a$, where $a$ is the lattice constant of Fe). The mesh of $128\times 128$ divisions of the full SBZ was used that resulted in 2145 $\mathbf{k}$-points in the irreducible wedge of the SBZ (ISBZ). (These 2145 $\mathbf{k}$ points of the ISBZ were used for plotting Fig 3.) For each transmission function corresponding  theoretical curve (shown by  red dashed line) that describes the small $|k|$ behavior of the transmission is also plotted. Theoretical curves for $T_{uu}(\mathbf{k})$ transmission were fitted according to the Eq. (\ref{T.1}) using $\gamma(0,E)$ shown on Fig. 2 (a) and two fitting constants: $A_{uu}$ and $\alpha$. Theoretical curves for $T_{ud}(\mathbf{k})$ transmission were fitted according to the Eq. (\ref{T.4}) for $E-E_F=-0.8$ eV and according to the Eq. (\ref{T.2}) for other energy points  with  additional fitting constant $f_{ud}$ that corresponds to the maximum value of the function $f_{ud}(\mathbf{k}/|k|)$, $f_{ud}=\max_{\mathbf{k}}f_{ud}(\mathbf{k}/|k|)$. Theoretical curves for  $T_{dd}(\mathbf{k})$ transmission were plotted according to the Eq. (\ref{T.5}) for $E-E_F=-0.8$ eV and according to the Eq. (\ref{T.3}) for other energy points \emph{without any additional fitting constants}.

\begin{figure}[h]
\includegraphics*[trim={0.5cm 9.6cm 0.2cm 1.2cm},clip,width=8.5cm]{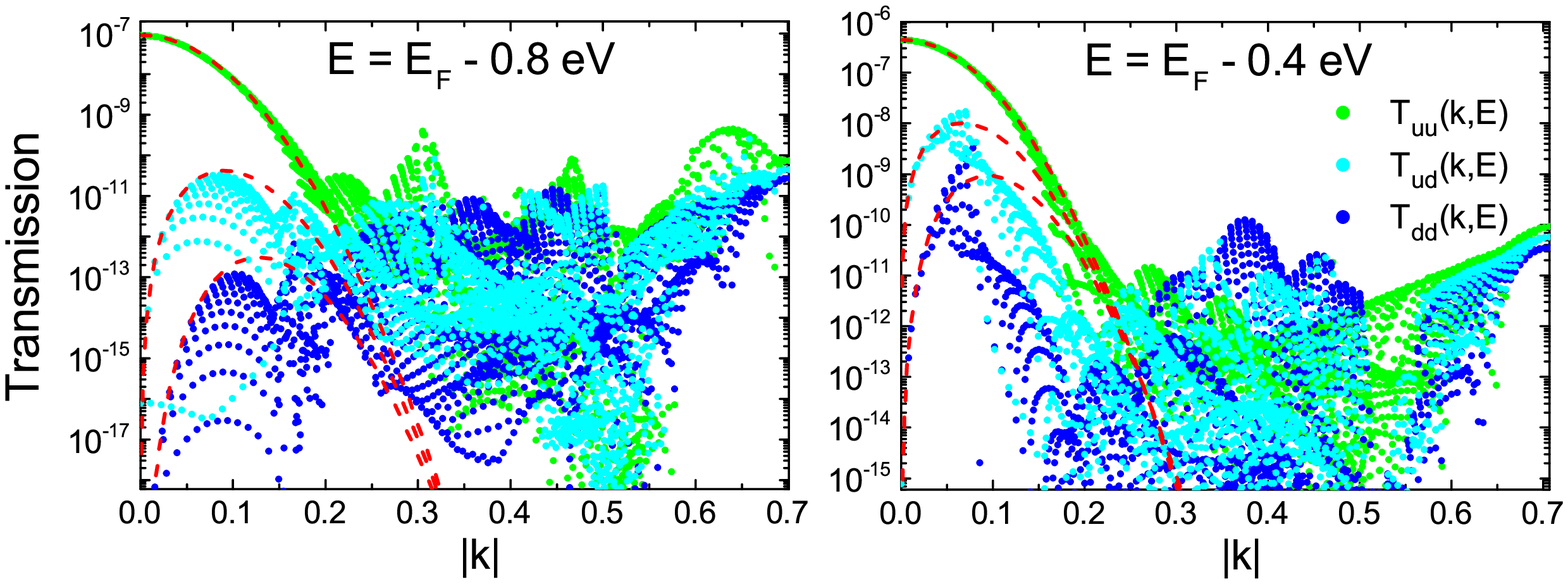}
\includegraphics*[trim={0.5cm 9.6cm 0.2cm 1.2cm},clip,width=8.5cm]{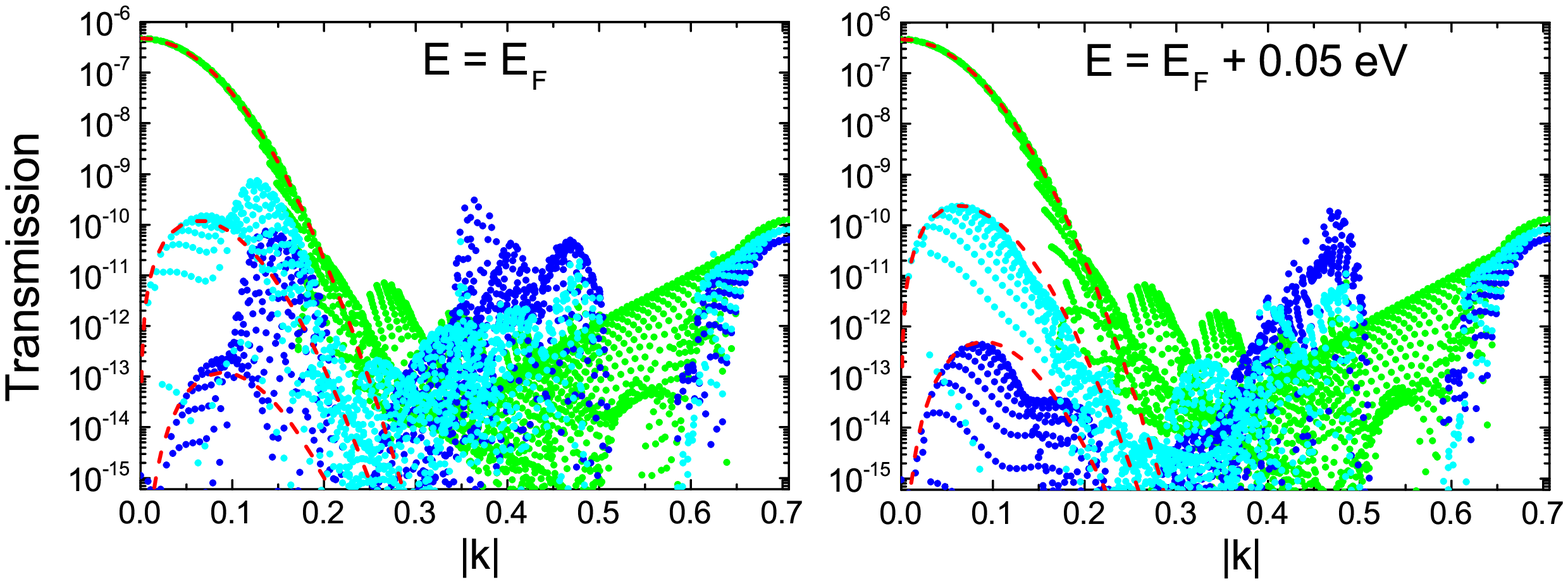}
\includegraphics*[trim={0.5cm 9.6cm 0.2cm 1.2cm},clip,width=8.5cm]{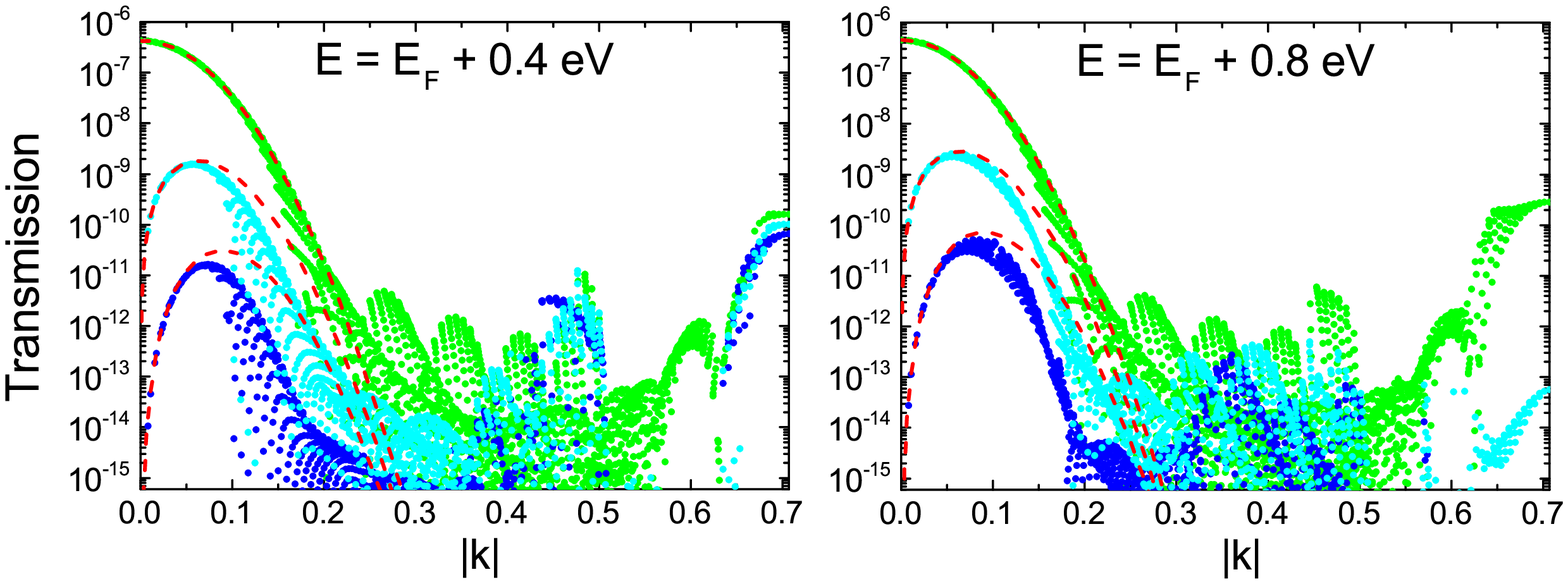}
\caption{(color online). Transmission functions $T_{uu}(\mathbf{k},E)$ (green dots), $T_{ud}(\mathbf{k},E)$ (cyan dots), and $T_{dd}(\mathbf{k},E)$ (blue dots) shown for the Fe/MgO/Fe MTJ with $N=10$ for 6 energies $E-E_F=-0.8, -0.4, 0, 0.05, 0.4$ and $0.8$ eV as function of the absolute value of the wave-vector $|k|$ (shown in units of $2\pi/a$). Red dashed lines are theoretical curves that describe behaviour of the transmission functions at small $|k|$. Theoretical curves were plotted using Eqs. (\ref{T.1},\ref{T.4}-\ref{T.5}) for $E-E_F=-0.8$ eV  and Eqs. (\ref{T.1}-\ref{T.3}) for all other energy points (see text for details).
}
\label{fig3}
\end{figure}

One can see that theoretical curves describe the small $|k|$ behavior of all transmission functions rather well  in   broad range of energies, including the $E-E_F=-0.8$ eV energy where behavior of the $T_{ud}(\mathbf{k})$ and $T_{dd}(\mathbf{k})$  changes from that described by Eqs. (\ref{T.2}-\ref{T.3}), to that described by Eqs. (\ref{T.4}-\ref{T.5}).  We  stress that behavior of the $T_{dd}(\mathbf{k})$ transmission is very well described by corresponding theoretical curve that was plotted without any additional fitting - by using only the constants derived from fitting the $T_{uu}(\mathbf{k})$ and $T_{ud}(\mathbf{k})$ functions (which provides yet another conformation of validity of Eq. (\ref{eqTT})).

For all six energy points theoretical curves correctly  predict small $|k|$ behavior of the $T_{uu}(\mathbf{k})$ function up to $|k|\sim 0.2$, where $T_{uu}(\mathbf{k})$ is reduced by many orders of magnitude from its maximum. Theoretical curves for $T_{ud}(\mathbf{k})$ and $T_{dd}(\mathbf{k})$ functions start to deviate from  actual transmissions at $|k|\sim 0.1$, where small $|k|$ approximation becomes invalid. The theoretical curves  correctly describe the local maximum of the $T_{ud}(\mathbf{k})$ for all considered energy points except $E-E_F=-0.4$ eV energy which is  a transitional point where $\Delta_5$ minority band disappears (see Fig 4 (c)).
Due to corresponding  Van Hove singularity in the density of states (DOS) at this energy the maximum of the $T_{ud}(\mathbf{k})$ is the largest one for $E-E_F=-0.4$ eV  as compared to maximums of $T_{ud}(\mathbf{k})$ for another five energy points (which leads to the smallest TMR at $N=10$ compared to other energy points, see Fig. 4 (a) and Fig. 5).

The global maximum of the function $T_{ud}(\mathbf{k})$ does not coincide with the local maximum described by the theoretical curves also for two other energy points: $E-E_F=-0.8$ eV and $E=E_F$. For $E-E_F=-0.8$ eV the small $|k|$ region is strongly suppressed by the $|k|^4$ factor (see Eq. \ref{T.4}), so $T_{ud}(\mathbf{k})$ near the $M$ point ($M$ point  on Fig. 3 corresponds to  largest $|k|=1/\sqrt{2}$)  is larger compared to $T_{ud}(\mathbf{k})$   near $|k|=0$. At sufficiently large $N$ the contribution from $|k|=0$ region will eventually become dominant, but this asymptotic has not been reached yet at $N=10$ for $E-E_F=-0.8$ eV.

The global maximum of the $T_{ud}(\mathbf{k})$ at the $E=E_F$ energy point reached at  $|k|\sim 0.15$ is not described by Eq. (\ref{T.2}) and corresponds to the interface resonance states existing in narrow energy window near $E_F$ \cite{Butler01,Belashchenko05,Rungger09,Tiusan04,Feng09,Faleev12}. The IRS are very sensitive to small changes of the energy and, as can be seen on Fig 3, the peak in $T_{ud}(\mathbf{k})$ associated with IRS disappears already at $E-E_F=0.05$ eV. The  IRS contribution to the APC transmission can be seen as narrow peak on Fig. 1(a) with maximum at $E-E_F=-0.009$ eV and width $\sim 0.02$ eV (at $N = 10$), and also, as narrow dip in the TMR, on Fig. 4(a).
We note that energy position of the IRS states is very sensitive to the details of the Fe/MgO interface and depends on the DFT functional used for relaxation of the interface structure \cite{Feng09}. In addition, recent beyond-DFT QSGW calculations show that IRS peak in DOS is shifted from $E=E_F$ (as predicted by DFT) to $E=E_F+0.12$ eV \cite{Faleev12}, which is in agreement with experimental measurements \cite{Zermatten08}.

\begin{figure}[h]
\includegraphics*[trim={0.6cm 1.8cm 3.6cm 2.0cm},clip,width=4.6cm]{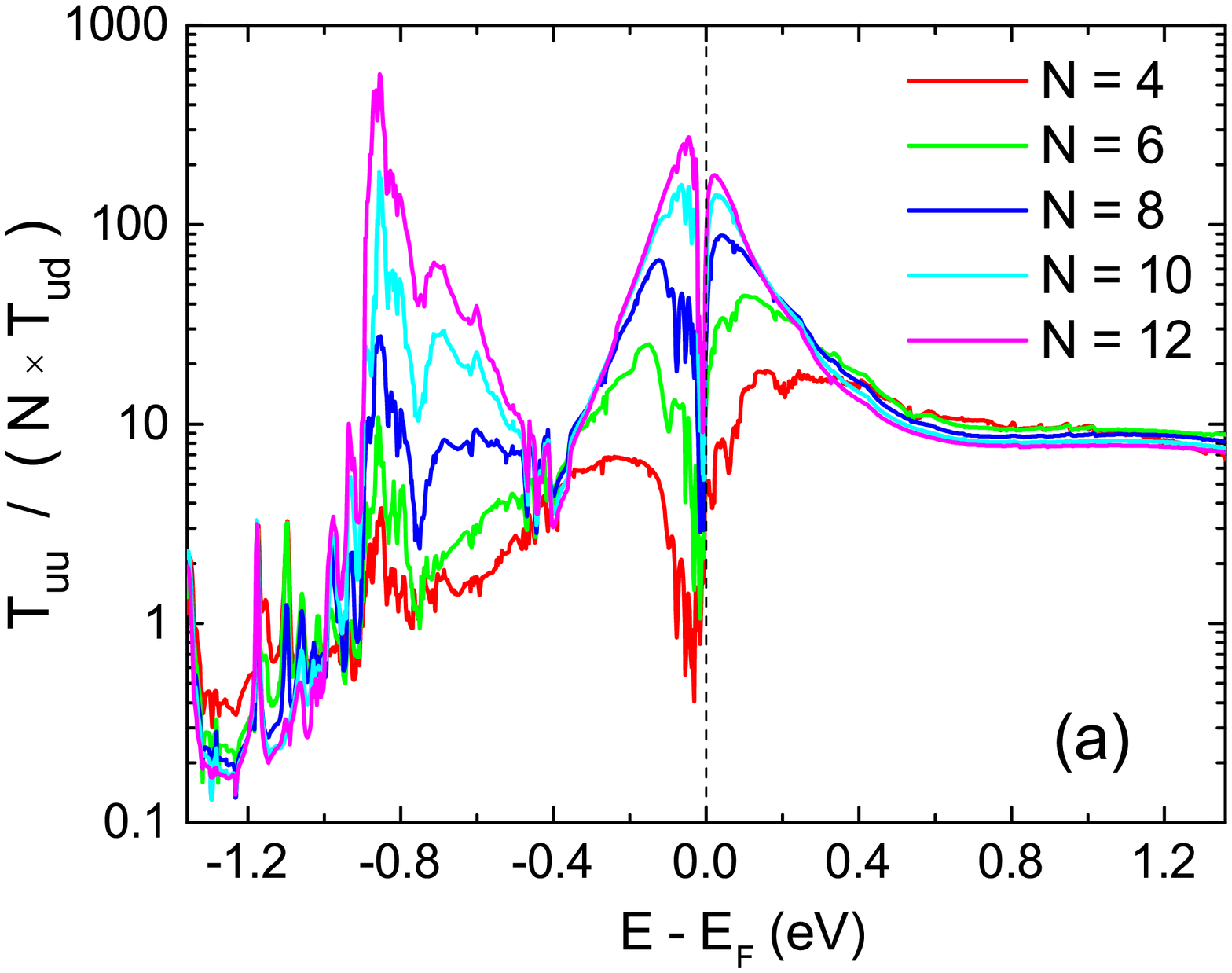}
\includegraphics*[trim={0.6cm 0.4cm 3.8cm 0.4cm},clip,width=4.0cm]{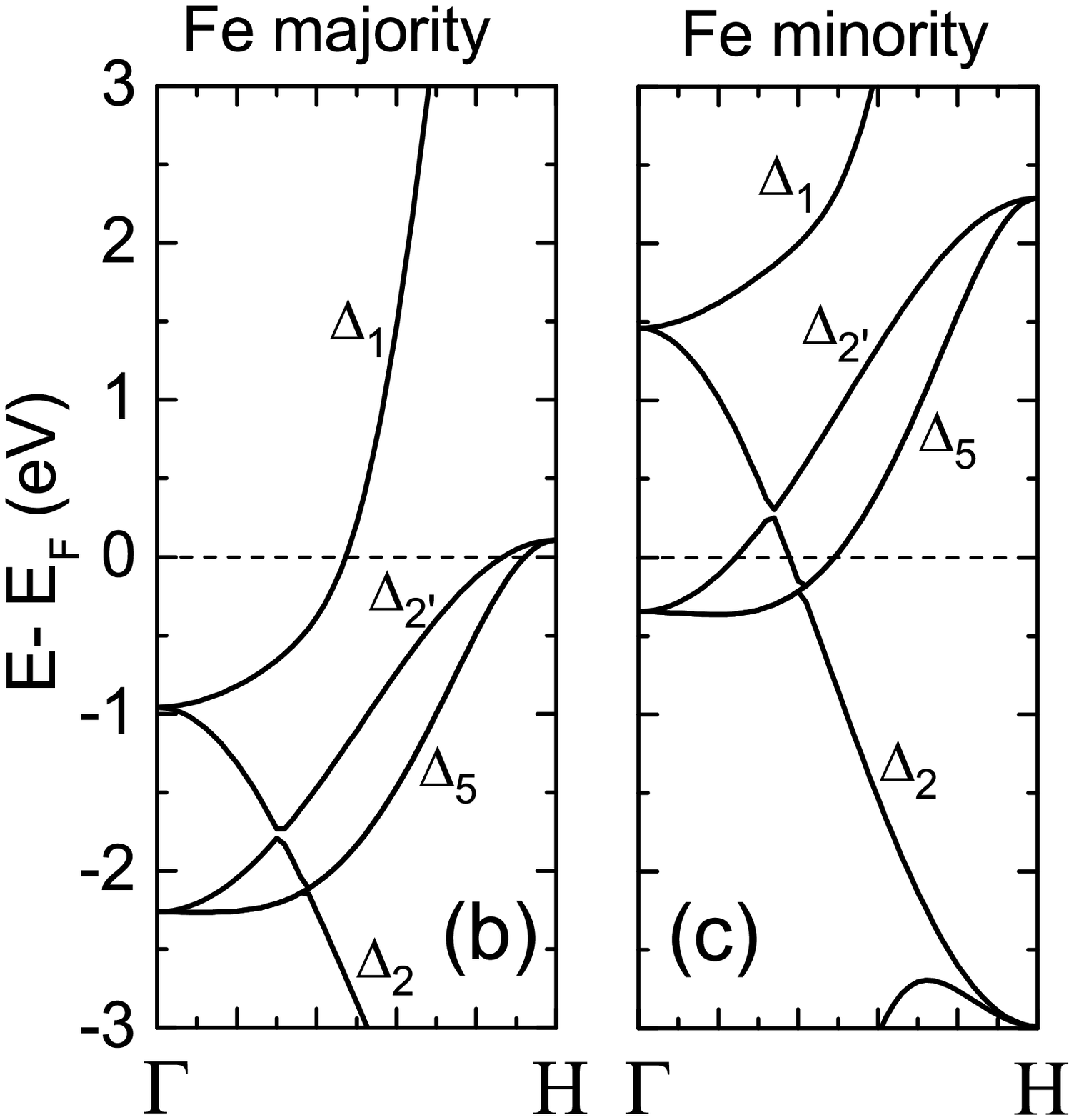}
\caption{(color online). (a) $T_{uu}(E)/(N \times T_{ud}(E))$ shown as function of the energy for $N=4,6,8,10$ and $12$. (b) majority and (c) minority Fe bands plotted along the $\Gamma-$H symmetry line.
}
\label{fig4}
\end{figure}
Figure 4(a) shows $T_{uu}(E)/(N \times T_{ud}(E))$ as function of the energy for $N=4,6,8,10$ and $12$. It is seen that  for all considered $N$ functions $T_{uu}(E)/(N \times T_{ud}(E))$ are very close to each other in energy range $E> E_F+0.4$ eV, thus confirming that linear with $N$ asymptotic behavior of the TMR (\ref{TMR}) resulted from \emph{native} symmetry filtering effect is established in Fe/MgO/Fe MTJ starting already with $N=4$  for $E> E_F+0.4$ eV. Established   linear asymptotical behavior is also seen on Fig. 5(b) where TMR is plotted as function of  $N$ for several energy points with $E \geqslant E_F+0.4$ eV.
For energies between, approximately, $E_F-0.2$ eV and $E_F+0.2$ eV the asymptotic behavior is reached at larger $N$ due to two factors (1) contribution of the IRS to $T_{ud}(E)$, and (2) generally smaller STF of minority electrons, $t_{d \mathbf{k} E}$, at $|k| \sim 0$ in this energy window (as compared, for example, to $t_{d \mathbf{k} E}$ at $E> E_F + 0.4$ eV) that leads to increased relative contribution to the $T_{ud}(E)$  from parts of the SBZ other then $|k|\sim 0$  (although the contribution of the $|k|\sim 0$ region to $T_{ud}(E)$ still increases with increased $N$).

As seen on Fig 4(a) and also on Fig 1(a), the width of the IRS peak reduces  with increased $N$ due to decaying of the IRS states with $|k| > 0$ inside the barrier with attenuation constant larger then $\gamma(0,E)$. As a result, curves with $N=10$ and $12$ shown on Fig 4(a) are very close to each other for the whole range $E>E_F-0.4$ eV, except small region with width $\sim 0.02$ eV near $E_F$ where the contribution of some IRS states (states with $|k| \sim 0$)  to the  $T_{ud}(E)$ still survives.

The fact that minority STF $t_{d \mathbf{k} E}$ at $|k| \sim 0$ for $E$ between $E_F-0.2$ eV and $E_F+0.2$ eV is smaller compared to that  outside of this energy window can also be seen by comparing the $\mathbf{k}$-resolved transmission $T_{ud}(\mathbf{k})$ on Fig. 2 for $E-E_F=0$ and $0.05$ eV with that for  $E-E_F=-0.4, 0.4$ and $0.8$ eV.  (Note that majority STF $t_{u \mathbf{k} E}$ does not change much in broad energy range $E>E_F-0.4$ eV, as can be concluded from comparing $T_{uu}(\mathbf{k})$ on panels corresponding to different energy points on Fig 2 and smooth behavior of the $T_{uu}(E)$ shown on Fig 1(b) and Fig 2(b).) Small STF of minority electrons for energies between $E_F-0.2$ eV and $E_F+0.2$ eV results in larger values of TMR for $N \geqslant 6$ at this energy window (see Fig. 4(a) and Fig. 5) as compared to the TMR outside of this window, but within broader window $E>E_F-0.4$ eV where $\Delta_5$ minority Fe state still exists. Finally, in the combination with the symmetry filtering effect,  reduced STF of minority electrons at the Fe/MgO interface at $|k| \sim 0$ is responsible for large $TMR>10,000\%$ predicted for Fe/MgO/Fe MTJ at $E=E_F$ for $N \geqslant 8$ in this and previous works.

In the energy window from $E_F-1.0$ eV to $E_F-0.4$ eV there is no $\Delta_5$-symmetry state along the $\Gamma-$H line in minority Fe channel (see Fig 4(c)), so $TMR \propto N$ asymptotic behavior changes to the $TMR \propto N^2$ asymptotic behavior (see Fig. 4(a) and Fig. 5(a,c)). The maximum of $T_{uu}(E)/(N \times T_{ud}(E))$ occurs at $E=E_F-0.85$ eV where $T_{ud}(E)$ is small due to the $|k|^4$ factor in $T_{ud}(\mathbf{k})$, while $T_{uu}(\mathbf{k})$ is enhanced due to the Van Hove singularity at the edge of the $\Delta_1$ majority Fe band (see Fig 1(b) and Fig 4(b)).

\begin{figure}[h]
\includegraphics*[trim={1.6cm 8.6cm 2.0cm 2.2cm},clip,width=8cm]{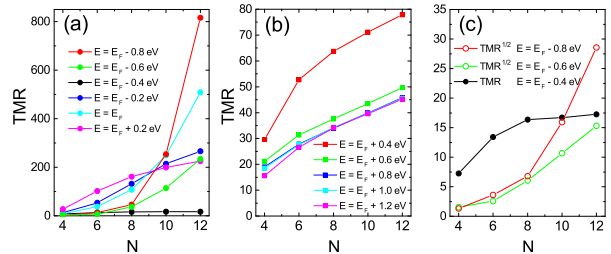}
\caption{ (color online).   Calculated TMR  shown as function of $N$ (a) for 6 energy points with $E \leqslant E_F+0.2$ eV and (b)  for 5 energy points with $E\geqslant E_F+0.4$ eV.  (c)  TMR for $E = E_F-0.4$ eV (shown on larger scale) and $TMR^{1/2}$ for $E - E_F=-0.8$ and $-0.6$ eV energy points.
}
\label{fig5}
\end{figure}
Calculated TMR is shown as function of $N$ on Fig 5(a) for 6 energy points with $E \leqslant E_F+0.2$ eV and on Fig 5(b) for 5 energy points with $E\geqslant E_F+0.4$ eV. Fig 5(c) shows TMR for $E = E_F-0.4$ eV (in the scale larger then that of Fig 5(a)) and $TMR^{1/2}$ for $E - E_F=-0.8$ and $-0.6$ eV.
The TMR shown on Fig 5 is calculated using the definition $TMR=T_{uu}/(2T_{ud}) -1$  that neglects the $T_{dd}$ contribution to transmission in PC. In general, $T_{dd}$  is much smaller compared to $T_{uu}$ except the case of small $N$ where at energies near $E_F$ IRS contribution to $T_{dd}$ is significant and $T_{dd}$ becomes comparable with (or even larger then) $T_{uu}$. As was noted in \cite{Belashchenko05} the contribution of IRS to $T_{dd}$ is significant due to the energy matching of the surface resonances at the left and right Fe/MgO surfaces that occurs "only for ideal, symmetric junctions, and only at zero bias". Slight non-ideality in any of the electrode or bias voltage as small as 0.01 V is sufficient to destroy this resonance matching \cite{Belashchenko05}.

The TMR curves shown on Fig 5 have linear with $N$ asymptotic behaviour for all energy points except $E-E_F=-0.8,-0.6 $ and $0$ eV. For energy points with $E\geqslant E_F+0.4$ eV  linear with $N$ behavior starts already with $N=4$. For $E-E_F=-0.4$ eV linear behavior starts somewhat latter, at $N=8$ due to approaching Fe minority $\Delta_5$ band edge  (see Fig. 4(c)) and corresponding reduction of the $\mathbf{k}$ integration range where $\psi^{\Delta_5}_d(\mathbf{k},k_z)$ bands still exist (see fast drop of the $T_{ud}(\mathbf{k},E)$) at $|k|\sim 0.08$ at this energy shown on Fig 2). For $E-E_F=-0.2$ and $0.2$ eV linear with $N$ behavior begins also somewhat latter, at $N=8$, due to generally small STF $t_{d \mathbf{k} E}$ at $|k|\sim 0$ for $E$ between $E_F-0.2$eV and $E_F+0.2$ eV and, therefore, enhanced weight of the contributions from other then $|k|\sim 0$ parts of the BZ at smaller $N$.

For $E=E_F$ linear asymptotic regime is not established yet even at $N=12$ due to narrow IRS-related peak in $T_{ud}(E)$ (although, as seen on Fig 4(a), linear asymptotic is established already for $N=10$ for energies just $0.1$ eV smaller or larger then $E_F$). (Here we stress again that in real experiment the contribution of the IRS will be suppressed due to interface roughness.)

As can be concluded from linear behavior of the $TMR^{1/2}$ as function of $N$ shown on Fig 5(c) for energy $E-E_F=-0.6$ eV, the  asymptotic behavior $TMR\propto N^2$   starts already with $N=6$. For $E-E_F=-0.8$ eV asymptotic behavior $TMR\propto N^2$  (or $TMR^{1/2}\propto N$) begins somewhat latter, at $N=8$, due to enhanced weight of the contributions from other then $|k|\sim 0$ parts of the BZ at smaller $N$, as we noted in discussion of the Fig 3.

In conclusion, from the analysis of the band structure of bulk Fe and complex band structure of MgO  we predicted \emph{native} to the symmetry spin filtering effect
asymptotical behavior of the TMR in Fe/MgO/Fe MTJ:
$TMR \propto N$   for energies from $E_F-0.4$ eV to $E_F+1.5$ eV and $TMR  \propto N^2$   for energies from $E_F-1.0$ eV to $E_F-0.4$ eV. \emph{Ab initio} calculations of transmission functions performed for the Fe/MgO/Fe MTJ confirm these theoretical predictions in broad range of energies and $N$.

Large TMR obtained  at energies near $E_F$  ($TMR> 10,000\%$ for $N\geqslant 8$) is attributed to the \emph{combination} of the symmetry spin filtering effect and small surface transmission function  of the minority Fe electrons at the Fe/MgO interface for $|k|\sim 0$ that leads to additional $\times 10$ enhancement of the $TMR$ at energy near  $E_F$ compared to that at $E>E_F+0.4$ eV or $E \sim E_F-0.4$ eV.

Super-linear behavior of $TMR$ at energies near $E_F$ obtained in this and previous theoretical works \cite{Butler01,Belashchenko05} is associated with contribution of the interfacial resonance states (quickly decaying with $N$) to the APC transmission. In real experiment the IRS contribution is suppressed due to  surface roughness thus providing natural explanation why no strong dependance of the TMR on $N$ have been found experimentally. Moreover, since the overlap integral at the Fe/MgO interface between Fe minority eigenstates and the $\Delta_1$-symmetry MgO eigenstate is proportional to $|k|^2$  at   $|k|\sim 0$ \emph{only} because of  mismatching symmetry of these eigenfunctions, surface roughness and/or interface chemical disorder that breaks the symmetry of the wave functions at the interface will inevitably lead to non-zero value of the the overlap integral at $|k|=0$ and therefore to saturation of the TMR at large $N$ which is observed experimentally \cite{Parkin04,Yuasa04}. In addition non ideal surface (due to interface chemical disorder or steps in surface layers) induces scattering of $|k|>0$ Fe minority states into $|k|=0$ MgO barrier eigenstate that  also leads to the saturation of the TMR at large $N$ \cite{Mathon06}.

Our prediction for the strength of the symmetry filtering effect is based on simple analysis of the band structure of bulk electrode material. Thus, such analysis could be used as a tool for quick material discovery search  of novel MTJs  in context of emerging technologies that requires high TMR, including STT-MRAM technology where the pool of candidate electrode materials includes several hundreds Heusler allays and magnetic multilayers.


S.F. and O.N.M acknowledge the CNMS User support by Oak Ridge National Laboratory
Division of Scientific User facilities. O.N.M acknowledge partial support  by C-SPIN, one of the six centers of STARnet,
a Semiconductor Research Corporation program, sponsored by MARCO and DARPA. S.F. would like to thank  Barbara Jones for useful discussions.


\begin{thebibliography}{99}

\bibitem{Butler01} W. H. Butler et al., Phys. Rev. B \textbf{63}, 054416 (2001).

\bibitem{Mathon01} J. Mathon and A. Umerski, Phys. Rev. B \textbf{63}, 220403(R) (2001).

\bibitem{Parkin04} S. S. P. Parkin et al., Nat. Mater. \textbf{3}, 862 (2004).

\bibitem{Yuasa04} S. Yuasa et al., Nat. Mater. 3, 868 (2004).

\bibitem{Belashchenko05} K. D. Belashchenko, J. Velev, and E. Y. Tsymbal,  Phys. Rev. B \textbf{72}, 140404(R) (2005)

\bibitem{Rungger09} I. Rungger, O. Mryasov, and S. Sanvito, Phys. Rev. B, \textbf{79}, 094414 (2009)

\bibitem{Tiusan04} C. Tiusam, J. Faure-Vinvent, C. Bellouard, et al., 
Phys. Rev. Lett. \textbf{93}, 106602 (2004).

\bibitem{Feng09} X. Feng, O. Bengone, M. Alouani, S. Lebegue, I. Rungger, and S. Sanvito, Phys. Rev. B \textbf{79}, 174414 (2009).

\bibitem{Faleev12} S. V. Faleev, O. N. Mryasov, and M. van Schilfgaarde, Phys. Rev. B, \textbf{85}, 174433 (2012)

\bibitem{Akerman05} Johan Akerman, Science, \textbf{308}, 508 (2005).

\bibitem{Belashchenko04} K. D. Belashchenko et al., Phys. Rev. B \textbf{69}, 174408 (2004)























\bibitem{Faleev05} S. V. Faleev, F. Leonard, D. A. Stewart, and M. van Schilfgaarde, Phys. Rev. B \textbf{71}, 195422 (2005).

\bibitem{Turek97} I. Turek, V. Drchal, J. Kudrnovsky, M. Sob, P. Weinberger, \emph{Electronic structure of disordered alloys, surfaces and interfaces}, (Kluwer, Boston, 1997)

\bibitem{Schilfgaarde98} M. van Schilfgaarde, W. R. L. Lambrecht, in \emph{Tight-binding approach to computational materials science}, edited by L. Colombo, A. Gonis, and P. Turchi, MRS Symposia Proceedings No. 491 
    (Pittsburgh, 1998).


\bibitem{Heiliger08} C. Heiliger, P. Zahn, B. Yu. Yavorsky, and I. Mertin, Phys. Rev. B \textbf{77}, 224407 (2008)

\bibitem{Mathon06} J. Mathon and A. Umerski, Phys. Rev. B \textbf{74}, 140404(R) (2006).

\bibitem{Worthmann04} D. Worthmann, G. Bihlmayer, and S. Blugel, J. Phys.: Cond. Matt. \textbf{16}, S5819 (2004).

\bibitem{Zermatten08} P.-J. Zermatten, et al., Phys. Rev. B \textbf{78}, 033301 (2008)














\end{thebibliography}
\end{document}